\documentclass{iopart}

\usepackage{iopams}
\usepackage{graphicx}

\begin{document}

\title{A hard disk analysis of momentum deficit due to dissipation}
\author{Ryoichi Kawai$^{1}$, Antoine Fruleux$^{2,3}$ and Ken Sekimoto$^{3,2}$}
\address{$^{1}$Department of Physics, University of Alabama at Birmingham, Birmingham, AL 35294, USA}
\address{$^{2}$Gulliver, CNRS-UMR7083, ESPCI, 75231 Paris, France}
\address{$^{3}$Mati\`{e}res et Syst\`{e}mes Complexes, CNRS-UMR7057, Universit\'e Paris-Diderot, 75205 Paris, France}

\begin{abstract}
When a Brownian object is in a nonequilibrium steady state, actual force exerted on it is different from one in a thermal equilibrium. In our previous paper [Phys. Rev. Lett. \textbf{108} (2012), 160601] we discovered a general principle which relates the missing force to dissipation rates through a concept of momentum deficit due to dissipation (MDD). In this article, we examine the principle using various models based on hard disk gases and Brownian pistons.  Explicit expressions of the forces are obtained analytically and the results are compared with molecular dynamics simulations.  The good agreement demonstrates the validity of MDD.
\end{abstract}

\submitto{\PS}
\maketitle

\section{Introduction}

Since its inception, understanding the effect of environments on a closed system has been a key subject of thermodynamics.  In particular when the system size is reduced to a mesoscopic scale, fluctuations in the system caused by those of the environments play dominant roles in many physical phenomena, such as Brownian motion.  The state of Brownian objects is typically investigated with the Langevin theory in which the environments exert forces on the Brownian objects through deterministic linear friction and fluctuating Langevin forces \cite{VanKampen}. This approach has been proven to be very effective in many applications.  Furthermore, the recent development of stochastic energetics \cite{Stochastic_Energetics} allows us to investigate the exchange of energy between the Brownian objects and the environments within the Langevin theory.

When the system is in contact with more than one environments which are not equilibrium with each other, 
we expect that energy and momentum flows between the system and the environments change in the way that the detailed balance is broken. Rigorously speaking, this loss of detailed balance brings the environments out of equilibrium at least in the vicinity of the system-environment interfaces. This aspect is not reflected in the Langevin description.
Consider, for example, cases where a Brownian object is simultaneously in contact with two different heat baths at different temperatures.  What force will be exerted on the Brownian object by the baths? A natural extension to the standard Langevin theory is to use the frictions and the stochastic forces from each bath assuming that the fluctuation-dissipation relation of second kind (i.e. the Einstein relation) 
holds independently for each bath, the condition which is sometimes called local detailed balance.
While such a simple linear model works well in many cases, there are phenomena that refuse to be understood by the linear Langevin approach. 

A most striking example is the {\it adiabatic piston} placed between two gases with different temperatures \cite{AdiabaticPiston-Piasecki,AdiabaticPiston-VdB}.  The piston has no internal degree of freedom so that no heat flows between the baths through it. An interesting question is if the piston moves when the two baths have the same pressure.  Naively one may think that the piston does not move because pressure on the both sides of the piston is the same. It turns out that the laws of thermodynamics alone cannot tell whether the piston moves or not \cite{Callen}. Feynman \cite{Feynman} pointed out that the fluctuations of the piston's velocity should be taken into account. However, the Langevin approach with the linear frictions falsely predicts zero mean velocity.

A similar difficulty also appears in some models of Brownian motors working between two baths \cite{BrownianMotors}.  In these models, the body of Brownian objects, e.~g. a triangular body, is not symmetric with respect to space inversion.  This kind of asymmetry is not fully realized in the linear Langevin equation since the linear friction constant or tensor is non-polar.  In the case of the adiabatic piston, the Brownian object itself is symmetric but the environments are not.  In either case, the linear Langevin theory cannot take into account asymmetric interactions between the Brownian objects and the environments.
                                                                                                                                                                                                                    
A common solution to these problems has been to resort to full and general microscopic descriptions, such as molecular dynamic (MD) simulation or master-Boltzmann equations under pertinent perturbative approximations. These methods are effective in predicting the outcome. For the adiabatic piston, the MD simulation \cite{AdiabaticPiston-VdB} and the perturbative master-Boltzmann equation  \cite{AdiabaticPiston-Piasecki,AdiabaticPiston-VdB} give quite consistent results showing that the piston moves towards the \textit{hotter} reservoir.  A more recent investigation based on the nonlinear Langevin equation also confirms it \cite{AdiabaticPiston-Plyukhin}.  While we now know that the piston moves, we still don't fully understand the physics behind it.

Recently, we investigated the force exerted by gas particles on a Brownian object in a non-equilibrium steady state (NESS) and discovered a rather general principle \cite{MDD}.  When there is energy dissipation, net momentum flux at the surface of a Brownian object is reduced from the flux without dissipation, which we shall call \textit{momentum deficiency due to dissipation} (MDD). As a consequence the force on the Brownian object decreases from the equilibrium force by
\begin{equation}
 F_\mathrm{MDD} = -c\, \frac{ J_{\mathrm{diss}}^{\mathrm{(e)}} }{v_\mathrm{th}}
\label{eq:main}
\end{equation}
where $J_{\mathrm{diss}}^{\mathrm{(e)}}$ is energy dissipation per unit time, and the thermal velocity of the gas particles of mass $m$ at temperature $T$ is defined by $v_\mathrm{th} \equiv \sqrt{k_\mathrm{B} T/m}$. The Boltzmann constant is denoted with $k_\mathrm{B}$.  The positive prefactor $c$  depends on the detail of the system but usually at the order of 1.

With this new principle, we are able to explain all of the above mentioned phenomena without the lengthy calculation \cite{MDD}. It is this MDD that is what is missing in the linear Langevin theory.  The stochastic energetics \cite{Stochastic_Energetics} tells us that the linear Langevin theory is sufficient to obtain the dissipation rate.  Therefore, one can evaluate the missing force (\ref{eq:main}) within the Langevin description. 
Furthermore, we note that this fundamental principle is applicable to systems beyond the regular Brownian objects, such as inelastic pistons and granular Brownian ratchets \cite{InelasticPiston,GranularRatchet}.

In this paper, we will demonstrate the validity of the new principle (\ref{eq:main}) using hard disk systems including MD simulation.  In the next section, we heuristically explain the idea of momentum deficit due to dissipation.  Then, we derive explicit expressions of $F_\mathrm{MDD}$ for hard disk systems.  The results are compared with MD simulation of shared Brownian pistons. We also derive non-equilibrium forces on an inelastic Brownian piston.  Despite that the origin of dissipation is quite different from the previous model, we arrive at the same expression, (\ref{eq:main}) and MD simulation confirms it.

\section{Momentum deficit due to dissipation: A heuristic argument}

\begin{figure}
\begin{center}
\includegraphics[width=8.0cm]{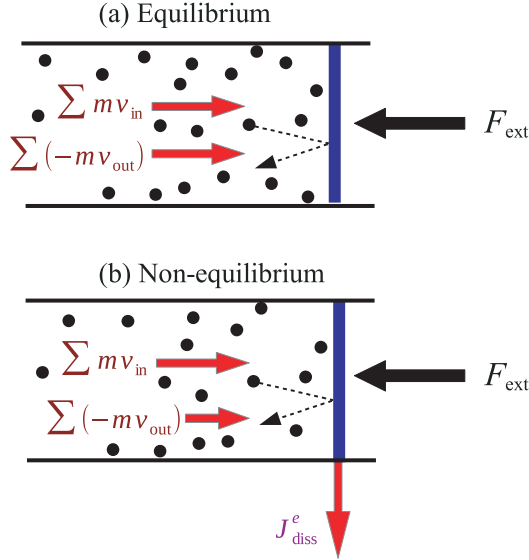}
\caption{Simple models illustrating the key point of MDD.  The upper panel (a) shows an equilibrium case where a gas is confined by a piston pressed with a constant force $F_\mathrm{ext}$. In this case, the momentum flow of the outgoing particles, $\sum(-m v_\mathrm{out})$ equals the momentum flow of the incoming particles $\sum m v_\mathrm{in}$.  The sum of the two flows is the pressure of the gas which balances with the external force.  The lower panel (b) illustrates a non-equilibrium case where the gas loses energy through the piston to external environments.  In this case, the momentum flow of the outgoing particles is smaller than that of the incoming particles, resulting in the reduction of the force exerted on the piston by the gas.}
\label{fig:key-idea}
\end{center}
\end{figure}

In this section, we briefly summarize the key concept developed in Ref. \cite{MDD}.  First, we consider a simple equilibrium system illustrated in Fig. \ref{fig:key-idea}(a).  The system consists of a two-dimensional cylinder filled with a gas and a piston of mass $M$ with surface size $L$, pressed with a constant external force $F_\mathrm{ext}$. The piston is a Brownian object and its velocity fluctuates due to the collision with the gas particles.  However, when the piston is in thermal equilibrium, its mean velocity is zero.   Hence, the sum of the pre-collisional and post-collisional momentum flows, $\sum m v_\mathrm{in}+ \sum \left (-m v_\mathrm{out}\right )$, is balanced by the external force.  Here the summation is taken over all collisions during a unit time.  The detailed balance tells us that at equilibrium the two momentum flows must be equal and therefore, on average, $2 \sum m v_\mathrm{in}  = \left |F_\mathrm{ext} \right | = p L$, where $p$ is the pressure of the gas. Note that unlike a simple kinetic theory used in elementary textbooks the individual collisions can transfer energy and momentum between the gas and the piston at the microscopic time scale since the piston is a Brownian object. It is the detailed balance that makes the two momentum flows identical on average.

Now we turn to a non-equilibrium case shown in Fig. \ref{fig:key-idea}(b) where energy flows from the gas through the piston into another environment.  We assume that the piston is in a NESS so that its mean velocity is zero. The magnitude of the external force is the same as that in the equilibrium case, i.~e., $F_\mathrm{ext}=p L$.  In non-equilibrium cases, this relation does not necessarily indicate mechanical equilibrium.  Even for a non-fluctuating macroscopic object, the actual force exerted by the gas deviates from $p L$ as observed in a radiometer\footnote{Maxwell struggled to explain it until near the end of his life.  See Ref. \cite{Radiometer}.}.  Our question is what is the actual force on the piston when it is a Brownian object.  We expect that the post-collisional speed is smaller than that in the equilibrium case.  Hence, the net momentum flow must be reduced. This is the momentum deficit due to dissipation and the force induced by MDD, $F_\mathrm{MDD}$, is defined by 
\begin{equation}
F_\mathrm{MDD} = \sum m v_\mathrm{in} + \sum (-m  v_\mathrm{out}) - p L\, ,
\end{equation}
which vanishes in the absence of dissipation.

The magnitude of $F_\mathrm{MDD}$ can be estimated from the energy balance
\begin{equation}
 \sum \frac{m}{2} v_\mathrm{in}^2 - \sum \frac{m}{2} v_\mathrm{out}^2 = J^\mathrm{(e)}_\mathrm{diss}\, .
\end{equation}
As a rough estimate, we replace fluctuating quantities with typical values; $\sum v_\mathrm{in} \sim \omega_\mathrm{col}\, v_\mathrm{th}$ and $\sum v_\mathrm{in}^2 \sim \omega_\mathrm{col}\, v_\mathrm{th}^2$ where $\omega_\mathrm{col}$ is the number of collisions per unit time.  Similarly for the outgoing particles, we introduce a typical velocity $\bar{v}_\mathrm{out}$.  Furthermore, the dissipation is assumed to be so weak that $v_\mathrm{in} - v_\mathrm{out} \approx 2 v_\mathrm{th}$.  Then, the balance of momentum and energy are expressed in simpler forms:
\begin{eqnarray}
&& F_\mathrm{MDD} \approx - \omega_\mathrm{col}\, m (v_\mathrm{th}+\bar{v}_\mathrm{out})   \\                                                                                                                                                                                                                                                                                       
&& J_\mathrm{diss}^\mathrm{(e)} \approx  \omega_\mathrm{col}\, m ( v_\mathrm{th}+\bar{v}_\mathrm{out} )\,  v_\mathrm{th} \, .
\end{eqnarray}
Eliminating the unknown quantity $\bar{v}_\mathrm{out}$, 
we obtain Eq. (\ref{eq:main}) except for the prefactor $c$ which is omitted in the above phenomenological argument since it depends on the system configuration.

Despite the drastically simple derivation, Eq. (\ref{eq:main}) agrees with the result of lengthy calculations except for the prefactor. 
Applying this principle we are able to explain the adiabatic piston and other phenomena both conceptually and quantitatively \cite{MDD}.

\section{Momentum deficiency due to dissipation in hard disk systems}

\subsection{Basic model}

\begin{figure}
\begin{center}
 \includegraphics[width=8.0cm]{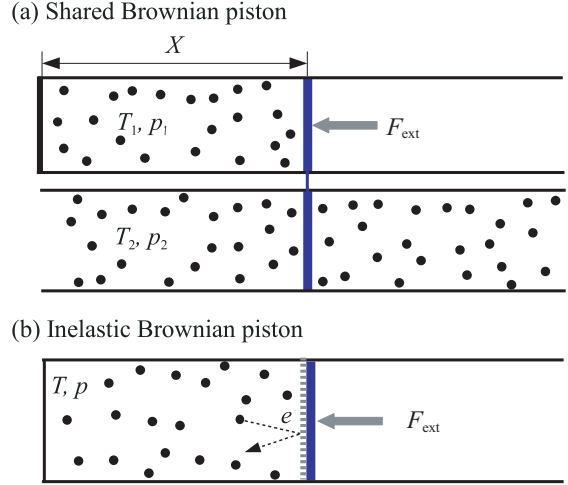}
 \caption{Models with two different types of dissipation.  (a) Shared Brownian piston: The piston is in contact with the second gas.  When  two gases have different temperatures, heat flows from the upper gas to the lower gas through the fluctuation of the piston.  (b) Inelastic Brownian piston: The collision between the gas particles and the piston is inelastic.  The energy dissipates into the internal degrees of freedome in the piston and the gas particles.}
\label{fig:model}
\end{center}
\end{figure}

We consider again the NESS case shown in Fig. \ref{fig:key-idea}(b).  In order to find explicit expressions we assume that the gas consists of hard disks which elastically collide with the piston.  The temperature of the gas, $T$, is assumed to be constant and the velocity of the gas particles satisfies the Maxwellian velocity distribution.\footnote{In general, this assumption does not hold under nonequilibrium conditions. Following the previous models \cite{AdiabaticPiston-VdB,BrownianMotors,InelasticPiston,GranularRatchet}, we assume that the incoming particles leave a thermostated region and directly hit the Brownian object.  The outgoing particles, lower in kinetic energy on average, travel back to the thermostated area before colliding the incoming particles, and thus the incoming particles follow the Maxwellian distribution.  However, this situation is possible only when the distance between the Brownian object and the thermostated region is not larger than the mean free path.}  We assume only binary hard collisions to take place.  The piston surface is smooth so that the velocity component parallel to the piston remains the same upon the collision. Hereafter, we consider only the velocity component perpendicular to the piston. Momentum and energy conserve at each collision  even when the Brownian object is simultaneously in contact with other baths or external agents since the hard disk collision is instantaneous. For $i$-th collision, the gas particle and the piston have the pre-collisional velocity $v_i$ and $V_i$, respectively and the corresponding post-collisional velocities $u_i$ and $U_i$ are determined by the momentum and energy conservation laws:
\begin{eqnarray}
&& m u_i - m v_i = M V_i - M U_i
\label{eq:p_con} 
\\
&& \frac{m}{2} u_i^2 - \frac{m}{2} v_i^2 = \frac{M}{2} V_i^2 - \frac{M}{2} U_i^2\, .
\label{eq:E_con}
\end{eqnarray}

Between the successive collisions, the piston interacts with another environment or an external agent and its momentum and energy change by $\Delta P_i$ and $\Delta E_i$, respectively.  The change in the piston velocity is determined by another set of momentum and energy balance equations:
\begin{eqnarray}
&& M V_{i+1} - M U_{i} = \Delta P_i
\label{eq:p_ext}
\\
&& \frac{M}{2} V_{i+1}^2 - \frac{M}{2} U_i^2 = \Delta E_i\, .
\label{eq:E_ext}
\end{eqnarray}
Summing up Eqs. (\ref{eq:p_con}) and (\ref{eq:p_ext}) over all $n$ collisions during a unit time, we find the net momentum balance:
\begin{equation}
\omega_\mathrm{col} \left ( m \langle v \rangle_\mathrm{col} -  m  \langle u \rangle_\mathrm{col} \right ) + F_\mathrm{ext} = 0\, ,
\end{equation}
where $F_\mathrm{ext} \equiv \sum \Delta P_i$ and we have used the NESS condition $M U_n=M V_1$.  The mean value is defined by $\sum v_i = \omega_\mathrm{col} \langle v \rangle_\mathrm{col}$.  Note that $\langle \cdots \rangle_\mathrm{col}$ indicates average over all collisions during a unit time (see Appendix) and it is not the same as a regular thermal average over the Maxwell distribution.  The force due to MDD is now expressed as
\begin{equation}
 F_\mathrm{MDD} = F_\mathrm{ext} - p L = - \omega_\mathrm{col} \, m \left (\langle v \rangle_\mathrm{col} + \langle u \rangle_\mathrm{col}  \right)\, .
\label{eq:F_disk1}
\end{equation}
Similarly, adding up Eq. (\ref{eq:E_con})  along with Eq. (\ref{eq:E_ext}) leads to net energy balance
\begin{equation}
J^\mathrm{(e)}_\mathrm{diss} = 
  \omega_\mathrm{col} \left ( \frac{m}{2} \langle u^2 \rangle_\mathrm{col}
-  \frac{m}{2} \langle v^2 \rangle_\mathrm{col} \right )\, ,
\label{eq:J_disk1}
\end{equation}
where $J^\mathrm{(e)}_\mathrm{diss} \equiv \sum \Delta E_i$. We have used the steady state condition
$M U_{n}^2/2 = M V_1^2 / 2 $.

For incoming particles, we find $\langle v \rangle_\mathrm{col} = \sqrt{\pi/2}\, v_\mathrm{th}$ and  $\langle v^2 \rangle_\mathrm{col}/\langle v \rangle^2_\mathrm{col} = 4/\pi$ (see Appendix).  We do not have exact statistics of the outgoing particles. However, when the dissipation is weak, we can assume that $\langle u \rangle_\mathrm{col}  \approx - \langle v \rangle_\mathrm{col} $ and $\langle v^2 \rangle_\mathrm{col} /\langle v \rangle^2_\mathrm{col}  \approx \langle u^2 \rangle_\mathrm{col} /\langle u \rangle_\mathrm{col} ^2 $.  Using this approximation, Eq. (\ref{eq:J_disk1})  is reduced to
\begin{equation}
  J_\mathrm{diss}^\mathrm{(e)} = \sqrt{\frac{8}{\pi}}\, v_\mathrm{th} \omega_\mathrm{col} \, m \left (\langle v \rangle_\mathrm{col}  + \langle u \rangle_\mathrm{col}  \right )\, .
\label{eq:J_disk2}
\end{equation}
Comparing Eqs. (\ref{eq:F_disk1}) and (\ref{eq:J_disk2}),  we obtain the formula (\ref{eq:main}) with a prefactor $c=\sqrt{\pi/8}$.

\subsection{Shared Brownian pistons}

Now, we introduce a more concrete NESS model shown in Fig. \ref{fig:model}(a) so that we can evaluate the dissipation rate.  The upper cylinder is the same as the basic model [Fig. \ref{fig:key-idea}(b)] and the gas in it has temperature $T_1$. A NESS condition is generated by linking the piston to another piston in the second cylinder filled with another gas at a different temperature $T_2$. These two pistons are rigidly connected and move together.   Unlike the upper one, the lower cylinder is periodic so that the particle density does not change as the piston moves.  Accordingly, when the piston moves, the pressure of the upper gas, $p_1$ changes while the pressure $p_2$ of the lower gas remains constant.  

It is known that when $T_1 > T_2$, heat flows from the upper gas to the lower gas through the fluctuations of the Brownian object \cite{HeatInRatchet}.
The heat dissipation through a shared piston is understood at the level of standard Langevin theory \cite{SharedPiston}.
Using the friction coefficients for the upper and lower pistons, $\gamma_1 = \sqrt{8/\pi} \rho_1 L \sqrt{k_\mathrm{B} T_1}$
and $\gamma_2=2 \sqrt{8/\pi} \rho_2 L \sqrt{k_\mathrm{B} T_2}$, we find the dissipation rate as
\begin{equation}
 J_\mathrm{diss}^\mathrm{(e)} = \sqrt{\frac{\pi}{8}}\, \frac{k_\mathrm{B} T_1 - k_\mathrm{B} T_2}{M(\gamma_1^{-1} + \gamma_2^{-1})}\, .
\end{equation}
Substituting this dissipation rate and the prefactor $c=\sqrt{\pi/8}$ to Eq. (\ref{eq:main}) , we obtain an explicit expression of the force due to MDD:
\begin{equation}
 F_\mathrm{MDD} = - \frac{2 \rho_1 \rho_2\, L}{\rho_1 + 2 \rho_2} \frac{m}{M} \left ( k_\mathrm{B} T_1 - k_\mathrm{B} T_2 \right )\, .
\label{eq:F_theory}
\end{equation}
\begin{figure}
\begin{center}
\includegraphics[width=8.0cm]{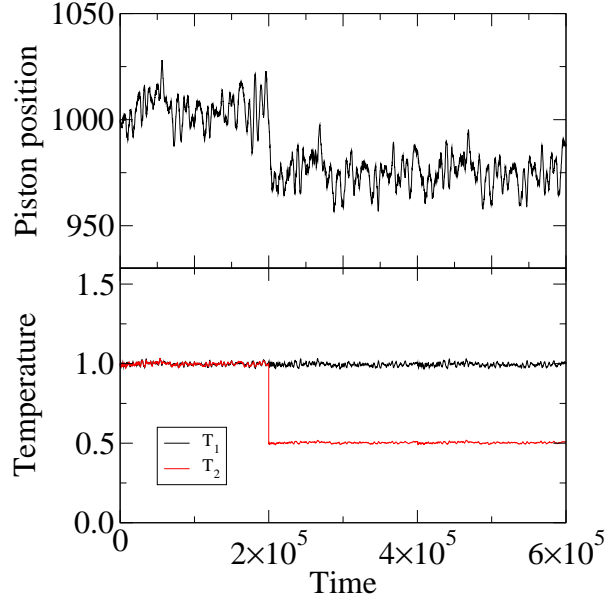}
\caption{Molecular dynamics simulation of momentum deficit due to dissipation using the model illustrated in Fig. \ref{fig:model}(a).  The upper panel shows the position of the piston and the lower panel the temperature of two gasses. Initially,the two gases have the same temperature $T_1=T_2=1.0$. When the temperature of the lower gas is reduced to $T_2=0.5$, the upper gas is compressed due to $F_\mathrm{MDD}$. The values of the system parameters are $L=300$, $M/m=20$, $X_0=1000$, $\rho_1=\rho_2=0.00333$ where the diameter of the gas particle is a unit of the length.  The mean free path in this configuration is about $150$.  The data are averaged over 300 realizations.}
\label{fig:traject}
\end{center}
\end{figure}

We have checked the above results using hard disk molecular dynamics simulation. The detailed simulation method will be written somewhere else.   Initially the system is at a thermal equilibrium with $T_1=T_2=1.0$. The mean position of the piston $X_0$ remains constant since $F_\mathrm{ext}+p_1 L=0$.  Figure \ref{fig:traject} shows that when the temperature of the lower gas is reduced to $T_2=0.5$,  the upper gas is compressed despite the temperature is kept at $T_1=1.0$. The displacement of the piston indicates that the force exerted on the piston by the gas is \textit{not} the pressure times the surface area.
Similarly, when $T_2$ is raised above $T_1$, the upper gas expands.  

When the system reaches a NESS, the piston is settled at a new position and a new pressure $p^\prime_1$ is established.    Assuming that the gas obeys the ideal gas law and the displacement $\Delta X$ is much smaller than $X_0$,  the missing force is estimated by
\begin{equation}
 F_\mathrm{MDD} = (p_1 - p_1^\prime) L = p_1 L\, \frac{\Delta X}{X_0}\, .
\label{eq:F_sim}
\end{equation}

\begin{figure}
\begin{center}
\includegraphics[width=8.0cm]{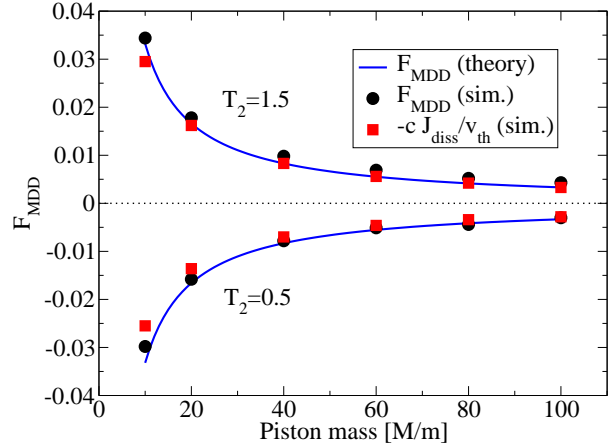} 
\caption{Molecular dynamics simulation of momentum deficit due to dissipation using the model illustrated in Fig. \ref{fig:model}(a).  The black solid circle shows the force measured in the simulation through Eq. (\ref{eq:F_sim}).  The red solid square plots the force estimated from Eq. (\ref{eq:main}) using the observed heat flow.  The full theoretical prediction Eq. (\ref{eq:F_theory}), is plotted with a solid line. 
The upper curves shows the case where the temperature of the second gas ($T_2=1.5$) is  higher than the upper gas whereas the lower curves shows the opposite case where the lower gas has $T_2=0.5$. See Fig. \ref{fig:traject} for the values of the system parameters.}
\label{fig:force}
\end{center}
\end{figure}

In Fig. \ref{fig:force} we plot $F_\mathrm{MDD}$ obtained in three different ways; Eq. (\ref{eq:F_sim}) with the measured displacement of the piston, Eq. (\ref{eq:main}) using the dissipation rate measured in the MD simulation, and the full theoretical result (\ref{eq:F_theory}). All three estimations agree very well, implying the validity of Eq. (\ref{eq:main}).

\section{MDD in granular systems: Inelastic piston}

\begin{figure}
 \begin{center}
  \includegraphics[width=8.0cm]{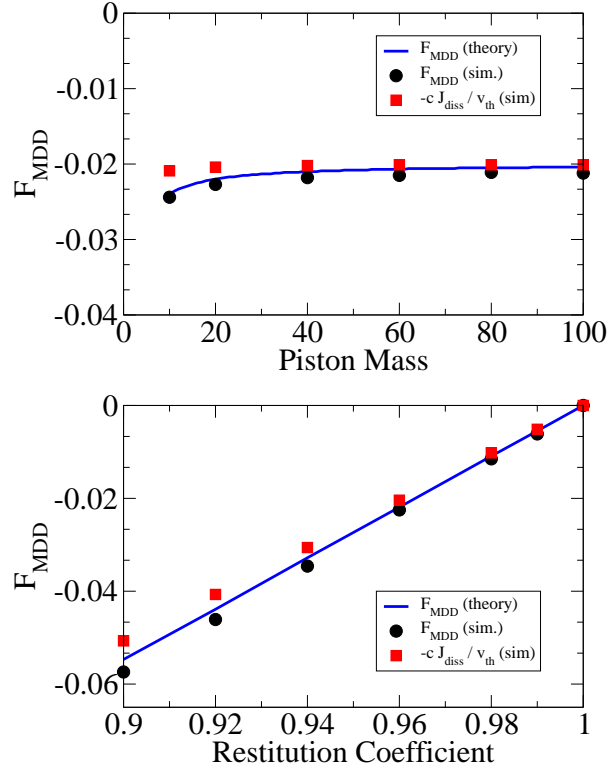}
  \caption{$F_\mathrm{MDD}$ on the inelastic piston shown in Fig. \ref{fig:model}(b) is plotted as a function of the piston mass (upper panel) and of the restitution coefficient (lower panel).  The force was evaluated in three different ways; the black circle indicates the force measured from the displacement of the piston in the MD simulation using Eq. (\ref{eq:F_sim}).
The red square shows the formula (\ref{eq:main}) using the dissipation rate observed in the MD simulation.  The solid line plots the theoretical value, the sum of Eqs. (\ref{eq:F_hk}) and  (\ref{eq:F_ex}). In the upper panel, $e=0.96$ and in the lower panel, $M/m=20$ are used. See Fig. \ref{fig:traject} for other parameter values.}
\label{fig:force_inelastic}
 \end{center}
\end{figure}

In order to demonstrate the generality of Eq. (\ref{eq:main}), we consider a different type of dissipation. In the model illustrated in Fig. \ref{fig:model}(b) energy dissipates into the internal degrees of freedom of the piston and gas particles though inelastic collisions between them. Using a standard collision rule used for granular systems, post-collisional velocities are related to the pre-collisional velocities as $u_i - U_i = -e (v_i-V_i)$ where $e$ is a coefficient of restitution ($0 < 1-e \ll 1$).  The momentum always conserves and thus Eq. (\ref{eq:p_con}) is still valid.  The energy balance for this model is
\begin{eqnarray}
\fl \frac{m}{2}v_i^2 -\frac{m}{2} u_i^2 &=& \frac{M}{2} U_i^2 - \frac{M}{2} V_i^2 
 +\frac{1-e^2}{2} \frac{m M}{m+M} (V_i-v_i)^2
\label{eq:E_loss}
\end{eqnarray}
Unlike the previous case, there is no dissipation between collisions, hence $V_{i+1}=U_i$.  Summing up Eq. (\ref{eq:E_loss}) for all collisions during a unit time, we obtained Eq. (\ref{eq:J_disk1})  again with a different dissipation rate: 
\begin{equation}
 J_\mathrm{diss}^\mathrm{(e)} \equiv =\frac{1-e^2}{2} \frac{m M}{m+M} \omega_\mathrm{col} \langle (V-v)^2 \rangle_\mathrm{col}
\label{eq:J_inel}
\end{equation}
Although the actual expression of $J_\mathrm{diss}^\mathrm{(e)}$ is different, the momentum and energy conservation laws are universal, and hence the general expression (\ref{eq:main}) is also valid for this model.

Now, we evaluate the dissipation rate (\ref{eq:J_inel}). Assuming that the piston obeys the Maxwellian distribution with a kinetic temperature $T_\mathrm{kin}$,  the mean value in Eq. (\ref{eq:J_inel}) is given by  
\begin{equation}
 \langle (v-V)^2 \rangle_\mathrm{col} = \frac{2 k_\mathrm{B} T}{m} + \frac{2 k_\mathrm{B} T_\mathrm{kin}}{M}\, ,
\end{equation}
as shown in Appendix. The dissipation rate is divided into two parts, one proportional to the mean kinetic energy of the gas particles (1st term) and the other to the mean kinetic energy of the piston (2nd term).  We shall call the former house keeping dissipation ($J_\mathrm{diss,hk}$) and the latter excess dissipation
($J_\mathrm{diss,ex}$) as coined in Ref.\cite{Oono}.  Using the mean values used in the previous section and $\omega_\mathrm{col} = \rho L v_\mathrm{th}/\sqrt{2 \pi}$ (see Appendix), we obtain
\begin{eqnarray}
&& J_\mathrm{diss,hk} =  
 (1-e)\sqrt{\frac{2}{\pi}}v_\mathrm{th} p L
\label{eq:J_hk}
\\
&& J_\mathrm{diss,ex} =  (1-e)\frac{\gamma}{M} \frac{k_\mathrm{B} T}{2}\, ,
\label{eq:J_ex}
\end{eqnarray}
where we assumed $m/M \ll 1$, $(1-e^2) \approx 2 (1-e)$, and $T_\mathrm{kin} \approx T \times (1+e)/2$ \cite{EffectiveT}.  Through our basic principle (\ref{eq:main}) and the prefactor $c=\sqrt{\pi/8}$, these dissipation rates lead to
$F_\mathrm{MDD} = F_\mathrm{MDD,hk}+F_\mathrm{MDD,ex}$ where
\begin{eqnarray}
&& F_\mathrm{MDD,hk} = -\frac{1}{2} (1-e) p L
\label{eq:F_hk}
\\
&& F_\mathrm{MDD,ex} = -\frac{m}{M} (1-e) p L\, .
\label{eq:F_ex}
\end{eqnarray}
Figure \ref{fig:force_inelastic} shows the result of MD simulation.  We again plot $F_\mathrm{MDD}$ evaluated in three different ways; Eq. (\ref{eq:F_sim}),
Eq. (\ref{eq:main}) using the measured dissipation rate, and full theoretical value with Eqs. (\ref{eq:F_hk}) and (\ref{eq:F_ex}). All three estimations agree well.

As discussed in Ref. \cite{MDD}, these forces explain the driving of inelastic pistons \cite{InelasticPiston} and granular ratchets \cite{GranularRatchet}. For the granular pistons, the house-keeping force (\ref{eq:F_hk}) is dominant and agrees with the perturbative results \cite{InelasticPiston}.  On the other hand, the driving force of the granular ratchets is the excess-dissipation force (\ref{eq:F_ex}) since the net house-keeping force vanishes in this model.

\section{Conclusions}

We examined the momentum deficit due to dissipation using various hard disk models.  The explicit expressions of the force due to MDD are obtained for two different models; the shared Brownian piston and the inelastic Brownian piston. Despite the different dissipation processes, the general principle (\ref{eq:main}) is valid for both cases.  Molecular dynamics simulations agreed well with the theoretical predictions for both cases.

\ack

We thank the organizers of NonEqSM2011 and European Science Foundation for providing us (K. S.  and R. K.) the opportunity to work on this project at NORDITA. R. K. also thanks the members of Physico-Chimie Th\'eorique at E.S.P.C.I. for their generous hospitality and stimulating discussion during his sabbatical stay.
K. S. thanks RIKEN for a financial support under the contract, CNRS: 30020830.

\setcounter{section}{1}
\appendix

\section{Collision statistics}

Here, we briefly explain the calculation of average over collision events.
We assume that the incoming  gas particles obey the Maxwell's velocity distribution $f_\mathrm{g}(v)=\sqrt{m/2\pi k_\mathrm{B} T} \mathrm{e}^{-m v^2/2 k_\mathrm{B} T}$ with temperature $T$.  Under non-equilibrium conditions, the velocity distribution of the piston is not necessarily Maxwellian. However, when the  piston is under a NESS not far from equilibrium, the piston approximately follows the Maxwell's velocity distribution $f_\mathrm{p}(V) = \sqrt{M/2\pi k_\mathrm{B} T_\mathrm{kin}} \mathrm{e}^{-M V^2/2 k_\mathrm{B} T_\mathrm{kin}}$ but with a kinetic temperature $T_\mathrm{kin}$ which is different from the temperature of the gas. Under these assumptions, the velocity distribution of gas particles colliding with the piston moving at velocity $V$ is given by
\begin{equation}
 \Phi(v;V) =  \frac{L \rho (v-V)}{\omega_\mathrm{col}} f_\mathrm{g}(v) f_\mathrm{p}(V) \Theta(v-V)
\label{eq:P_col}
\end{equation}
where $\Theta(\cdot)$ is the Heaviside step function and the normalization constant $\omega_\mathrm{col}$ is the total number of collisions per unit time defined by 
\begin{eqnarray}
  \omega_\mathrm{col} &=&  L \rho \int_{-\infty}^{\infty} \mathrm{d}V \int_V^{\infty} \mathrm{d}v (v-V) f_\mathrm{g}(v) f_\mathrm{p}(V) \nonumber\\
&=& L \rho \sqrt{\frac{k_\mathrm{B} T}{2 \pi m}} \sqrt{1+\frac{ m T_\mathrm{kin}}{M T}} \approx L \rho \frac{ v_\mathrm{th}}{\sqrt{2\pi}}
\end{eqnarray}
where $m \ll M$ is assumed.

Using the probability distribution (\ref{eq:P_col}) we obtain the 1st and 2nd moments:
\begin{eqnarray}
\fl \langle v \rangle_\mathrm{col} =  \int_{-\infty}^{\infty} \mathrm{d}V \int_{-\infty}^{\infty} \mathrm{d}v\, v\, \Phi(v;V)  \nonumber\\
= \sqrt{\frac{\pi k_\mathrm{B} T}{2 m}} \left ( 1+\frac{ m T_\mathrm{kin}}{M T}\right )^{-\frac{1}{2}} \approx \sqrt{\frac{\pi}{2}} v_\mathrm{th}
\\
\fl \langle v^2 \rangle_\mathrm{col} =   \int_{-\infty}^{\infty} \mathrm{d}V \int_{-\infty}^{\infty} \mathrm{d}v\, v^2\, \Phi(v;V)  \nonumber\\
= \frac{k_\mathrm{B} T}{m} \cdot \frac{m T_\mathrm{kin} + 2 M T}{m T_\mathrm{kin} + M T} \approx 2 v_\mathrm{th}^2
\end{eqnarray}
Similarly, the second moment of the relative velocity is computed as
\begin{eqnarray}
\fl \langle (v-V)^2 \rangle_\mathrm{col} =   \int_{-\infty}^{\infty} \mathrm{d}V \int_{-\infty}^{\infty} \mathrm{d}v\, (v-V)^2\, \Phi(v;V)  \nonumber\\
= \frac{2 k_\mathrm{B} T}{m} + \frac{2 k_\mathrm{B} T_\mathrm{kin}}{M} 
\end{eqnarray}
which is exact under the present assumption.

\end{document}